\documentclass[12pt]{article}

\input epsf
\usepackage{amssymb,amsmath,subfigure,float}
\usepackage[matrix,arrow,curve]{xy}
\usepackage{cite}
\usepackage{multirow}
\usepackage{graphicx,color}
\usepackage[debug,pageanchor=false]{hyperref}
\definecolor{link}{rgb}{.8,.15,.1}
\hypersetup{colorlinks=true,linkcolor=link,citecolor=link,urlcolor=link,linktocpage}

\makeatletter
\@addtoreset{equation}{section}
\makeatother

\DeclareMathOperator{\I}{Im}
\newcommand{\ee}{\mathrm{e}}
\def\squad{\,\,\,}
\newcommand{\cG}{G}
\newcommand{\im}{\mathrm{i}}
\newcommand{\cV}{\mathcal{V}}
\newcommand{\cB}{\mathcal{B}}
\newcommand{\Iprod}[2]{\langle {#1}, {#2} \rangle}
\newcommand{\nv}{n_\mathrm{v}}

\begin{document}

\begin{titlepage}

\begin{center}

\vskip .3in \noindent

{\Large \bf{Wilson lines for AdS$_5$ black strings\\\vspace{.2cm}}}

\bigskip

	Kiril Hristov and Stefanos Katmadas\\

       \bigskip
		 Dipartimento di Fisica, Universit\`a di Milano--Bicocca, I-20126 Milano, Italy\\
       and\\
       INFN, sezione di Milano--Bicocca,
       I-20126 Milano, Italy

       \vskip .5in
           
       {\bf Abstract }
        \vskip .1in        
        \end{center}         
       
We describe a simple method of extending AdS$_5$ black string solutions of 5d gauged supergravity in a supersymmetric way by addition of Wilson lines along a circular direction in space. When this direction is chosen along the string, and due to the specific form of 5d supergravity that features Chern-Simons terms, the existence of magnetic charges automatically generates conserved electric charges in a 5d analogue of the Witten effect. Therefore we find a rather generic, model-independent way of adding electric charges to already existing solutions with no backreaction from the geometry or breaking of any symmetry. We use this method to explicitly write down more general versions of the Benini-Bobev black strings \cite{Benini:2013cda} and comment on the implications for the dual field theory and the similarities with generalizations of the Cacciatori-Klemm black holes \cite{Cacciatori:2009iz} in AdS$_4$.	
       

\noindent

\vfill
\eject

\end{titlepage}

\section{Main method} 
\label{sec:intro}
In this short note we present a method of adding Wilson lines to already existing 5d black string solutions in a 
supersymmetry preserving way. It is a well-known feature of black string and black ring solutions in ungauged supergravity
\cite{Bena:2004wv, deWit:2009de, Gauntlett:2004qy}
that Wilson lines are zero-modes of the equations of motion and supersymmetry variations, which
lead to nontrivial electric charges in the full asymptotically flat solution. This fact seems to have been overlooked
in the literature dealing with asymptotically AdS solutions in 5d gauged supergravity
\cite{Benini:2013cda,Klemm:2000nj,Cucu:2003bm,Cucu:2003yk,Naka:2002jz,Gauntlett:2006qw,Almuhairi:2011ws,Hristov:2013xza}.
Here we aim to close this gap and show that, apart from a few subtleties, one can extend the known black string solutions
in the same way. This amounts to introducing a non-vanishing background gauge fields
\begin{equation}\label{wilson}
 A^I_{\gamma} = w^I\ ,
\end{equation}
where the direction $\gamma$ is along an isometry of the solution and the index $I$ runs over the number of gauge fields.
We will focus on black string solutions supported by magnetic charges, whose horizons can be Riemann surfaces of any genus.
One obvious choice for $\gamma$ is the direction along the black string, a circular direction in space (if we
foliate the asymptotic AdS$_5$ appropriately). An alternative choice\footnote{We thank Nikolay Bobev for pointing out this possibility.} arises for black strings with a horizon of genus $g > 0$, for which the direction $\gamma$ can be along
the Riemann surface throughout spacetime, since there exist nontrivial one-cycles in these cases (assuming again an
appropriate foliation of AdS$_5$). In both cases the Wilson lines \eqref{wilson} cannot be gauged away, but only periodically identified,
as $w^I \sim w^I + 2 \pi n$, where $n\in \mathbb{Z}$.

The addition of these Wilson lines, trivial as it might seem, actually introduces interesting features of the black string
solutions. When the direction $\gamma$ is taken to be along the string, the Wilson lines lead to nonvanishing conserved
electric charges in the spacetime due to the Chern-Simons terms present in 5d supergravity and the fact that black strings
are already supported by magnetic charge. The resulting electric charges are therefore proportional to both the Wilson lines
and the magnetic charges, similarly to the Witten effect \cite{Witten:1979ey} in 4d that requires a nonvanishing theta-angle.
In fact, as can be seen explicitly by dimensional reduction \cite{BarischDick:2012gj, Hristov:2014eza}, the Wilson lines can
be thought of as a 5d analogue of the theta-angle. On the other hand, if the direction $\gamma$ in \eqref{wilson} is taken
along the Riemann surface, one obtains a deformation of the original magnetically charged solution (c.f.\ \cite{Bah:2012dg} for analogous results in 7d) with no electric charges but an interesting dual field theory interpretation discussed in section \ref{sec:cft}.

Consider the action of a general $N=2$ $D=5$ Fayet-Iliopoulos (FI) gauged supergravity with an arbitrary number $n_v$ of vector 
multiplets\footnote{One can naturally extend the following arguments to more general theories with hypermultiplets with 
virtually no difference.}. The bosonic fields are the metric $g_{\mu \nu}$, $n_v$ real scalars $\phi^{i}$, and $n_v+1$ 
$U(1)$ gauge fields $A^I_{\mu}$. The bosonic Lagrangian in standard conventions \cite{Klemm:2000nj} is given by
\begin{align}
e^{-1}{\cal {L}} =&\, \frac{1}{2}R+g^{2}V
-{\frac{1}{4}}G_{IJ}F_{\mu \nu}{}^{I}F^{\mu \nu J}
-\frac{1}{2}{\cal G}_{ij}\partial _{\mu }\phi^{i}\partial ^{\mu }\phi ^{j} 
\nonumber \\
&\,+\frac{e^{-1}}{48}\epsilon^{\mu \nu \rho \sigma \lambda }C_{IJK}F_{\mu
\nu }^{I}F_{\rho \sigma }^{J}A_{\lambda }^{K}\,, 
\label{lagr}
\end{align}
with a gauge coupling constant $g$ and a scalar potential depending on the constant FI parameters $V_I$ and the $n_v+1$ sections, $X^{I} (\phi^i)$,
which parameterise the physical scalar fields when the condition 
\begin{equation}
{\cal V}={\frac{1}{6}}C_{IJK}X^{I}X^{J}X^{K}=1 \,,
\end{equation}
is satisfied. The Lagrangian is completely specified by the constant symmetric tensor of coefficients $C_{I J K}$.
All the physical quantities in \eqref{lagr} can be expressed in terms of the homogeneous cubic polynomial ${\cal V}$,
i.e.\ one can uniquely determine the scalar and gauge field kinetic terms ${\cal G}_{ij} (\phi)$ and $G_{I J} (\phi)$
from the coefficients $C_{I J K}$. 

Let us now consider a black string that is already satisfying the equations of motion. In the known examples in
gauged supergravity \cite{Benini:2013cda,Klemm:2000nj,Cucu:2003bm,Cucu:2003yk,Naka:2002jz,Gauntlett:2006qw,Almuhairi:2011ws,Hristov:2013xza} (and similar to the ones in ungauged supergravity \cite{Bena:2004wv, deWit:2009de, Gauntlett:2004qy}), the metric and field strengths are
of the form
\begin{equation}
 {\rm d} s^2 = f(r)^2\ \left( -{\rm d} t^2 + {\rm d} r^2 + {\rm d} z^2 \right) + h(r)^2\ {\rm d} \sigma^2_{\Sigma}\ 
,
\end{equation}
\begin{equation}
 F^I = p^I\ Vol_{\Sigma}\ ,
\end{equation}
and scalars $\phi^i(r)$. The functions $f(r), h(r)$ are typically known only numerically and in the limiting cases 
correspond to a metric interpolating between AdS$_3 \times \Sigma^2$ near the horizon and asymptotically locally AdS$_5$ 
foliated in $\mathbb{R}_t \times$S$^1 \times \Sigma^2$ coordinates. 

Now we want to argue that the addition of the Wilson lines \eqref{wilson} to the gauge field solution above immediately 
solves the equations of motion without changing any other detail of the solution (no matter whether direction $\gamma$ is chosen to be inside $\Sigma^2$ or to correspond to $z$). First note that the field strengths do 
not change upon the addition of Wilson lines, and bare gauge fields only enter in the Lagrangian \eqref{lagr} via the 
Chern-Simons term 
$$\epsilon ^{\mu \nu \rho \sigma \lambda }C_{IJK}F_{\mu \nu }^{I}F_{\rho \sigma }^{J}A_{\lambda }^{K}\ .$$
This term does not couple with either the metric or scalars, therefore one just needs to make sure that the Maxwell 
equations are satisfied to conclude that we have found a new solution. But even this task is trivial since the Maxwell 
equations no longer depend on the bare gauge fields even if the Lagrangian does. 

Let us now look at supersymmetry - assuming that the starting black string did lead to vanishing variations, we just 
have to make sure there are no new contributions from the Wilson lines. The relevant gravitino and gaugino variations are
\begin{eqnarray}
\delta \psi _{\mu } &=&\left( {\cal D}_{\mu }+\frac{i}{8}X_{I}(\Gamma _{\mu
}{}^{\nu \rho }-4\delta _{\mu }{}^{\nu }\Gamma ^{\rho })F_{\nu \rho }{}^{I}+%
\frac{1}{2}g\Gamma _{\mu }X^{I}V_{I}-\frac{3i}{2}gV_{I}A_{\mu }^{I}\right)
\epsilon ,  \label{gravitino} \\
\delta \lambda _{i} &=&\left( {\frac{3}{8}}\Gamma ^{\mu \nu }F_{\mu \nu
}^{I}\partial _{i}X_{I}-{\frac{i}{2}}{\cal G}_{ij}\Gamma ^{\mu }\partial
_{\mu }\phi ^{j}+\frac{3i}{2}gV_{I}\partial _{i}X^{I}\right) \epsilon ,
\label{gaugino}
\end{eqnarray}
where we used $X_{I}\!\equiv\!{\frac{1}{6}}C_{IJK}X^{J}X^{K}$ for brevity, $\epsilon$ is the supersymmetry parameter and
${\cal {D}_{\mu }}$ is the covariant derivative. The gaugino variation \eqref{gaugino} again does not depend on the
bare gauge fields and is therefore immediately zero, while the only new term in the $\gamma$-component gravitino
variation \eqref{gravitino} leads to the condition 
\begin{equation}\label{WilsonBPS}
 g\, V_I w^I = 0\ .
\end{equation}
Note that in ungauged supergravity we have $g=0$ and therefore BPS-ness is guaranteed automatically. Here instead we 
need to require that the linear combination $V_I w^I$ vanishes to guarantee supersymmetry.

We have thus shown that the addition of the $n_v+1$ Wilson lines \eqref{wilson} leads to a new black string solution and 
furthermore that supersymmetry restricts one linear combination of them through \eqref{WilsonBPS}, adding $n_v$
new parameters to the family of magnetic BPS solutions. When the Wilson lines are chosen to be along the Riemann surface ($\gamma \in \Sigma^2$),
one finds an extended class of black string solutions, with the same charges as before and some additional parameters.

However, if the Wilson lines are along the string direction ($\gamma \equiv z$) these additional parameters actually correspond to
electric charges due to the Chern-Simons term. The Chern-Simons term contributes
to the Maxwell equations
\begin{equation}
 \partial_{\nu}  (G_{I J} F^{\mu \nu J}) = \frac18\ e^{-1} \epsilon^{\mu \nu \rho \sigma \lambda }C_{IJK}F_{
\nu \rho}^{J}F_{\sigma \lambda}^{K}\ ,
\end{equation}
which lead to the following conserved electric charges
\begin{equation}\label{elcharges}
  q^I = -\ C_{I J K} w^J p^K\ .
\end{equation}
The new BPS black strings generated from the Wilson lines therefore have $n_v$ independent electric charges.

It is important to note that the procedure described above is the simplest one that allows to add Wilson lines,
since they are constant throughout spacetime. In analogous examples in ungauged supergravity
\cite{Bena:2004wv, deWit:2009de, Gauntlett:2004qy}, the most general solution allows for extra functions that
reduce to constant Wilson lines at the attractor. It is an interesting problem to consider the existence of such
solutions in gauged supergravity as they would correspond to more general electrically charged black
strings.

\section{Extending Benini-Bobev solutions} 
\label{sec:bb}
Following the above procedure, it is straightforward to write down a generalization of the Benini-Bobev (BB) black 
strings \cite{Benini:2013cda} that put together earlier solutions \cite{Klemm:2000nj,Cucu:2003bm,Cucu:2003yk,Naka:2002jz,Gauntlett:2006qw,Almuhairi:2011ws}. Consider the supergravity \eqref{lagr} with two vector multiplets (the so-called STU 
model), which is a truncation of the maximal $N=8$ supergravity arising from the compactification of type IIB 
supergravity on S$^5$. We have $V_1 = V_2 = V_3 = \frac{1}{3}$ and $C_{1 2 3} = 1$ and its permutations as only 
nonvanishing components. The bosonic fields in the solution, already adding the Wilson lines, are given by:
\begin{equation}
 {\rm d} s^2 = e^{2 f(r)}\ (-{\rm d} t^2 + {\rm d} z^2 + {\rm d} r^2) + e^{2 g(r)}\ {\rm d} \sigma^2_{\Sigma}\ ,
\end{equation}
\begin{equation}\label{gaugefields}
 F^I = - p^I\ Vol_{\Sigma}, \qquad A^I_{\gamma} = w^I\ ,
\end{equation}
\begin{equation}
 X^1 = e^{-\frac{\phi^1 (r)}{\sqrt{6}}-\frac{\phi^2 (r)}{\sqrt{2}}}, \quad X^2 = e^{-\frac{\phi^1 
(r)}{\sqrt{6}}+\frac{\phi^2 (r)}{\sqrt{2}}}, \quad X^3 = e^{ \frac{2 \phi^1 (r)}{\sqrt{6}}}\ .
\end{equation}
The full flow is given in terms of the functions $f(r), g(r), \phi^{1,2} (r)$ that were found numerically in 
\cite{Benini:2013cda} and are fully determined by the magnetic charges $p^I$ (or $a_I$ in the notation of \cite{Benini:2013cda}). These satisfy a further constraint imposed 
by supersymmetry, 
\begin{equation}\label{dirac}
 p^1+p^2+p^3 = - \kappa\ , 
\end{equation}
with $\kappa= +1, -1,$ or $0$ depending on the curvature of the Riemann surface $\Sigma$. The extra BPS condition \eqref{WilsonBPS} for the Wilson 
lines now becomes
\begin{equation}
 w^1+w^2+w^3 = 0\ . 
\end{equation}
There are conserved electric charges when $\gamma \equiv z$, which are found from \eqref{elcharges}
\begin{equation}
 q_1 = -(p^2 w^3 + p^3 w^2), \quad q_2 = -(p^1 w^3 + p^3 w^1), \quad q_3 = -(p^1 w^2 + p^2 w^1)\ .
\end{equation}

\subsection{Near-horizon BTZ with Wilson lines} 
\label{subsec:btz}
Near the horizon we can do better and write down a full analytic solution, which has the metric in the form BTZ$\times 
\Sigma^2$:
\begin{equation}
 {\rm d} s^2 = R^2_{AdS_3}\ {\rm d} s^2_{BTZ} + R^2_{\Sigma}\ {\rm d} \sigma^2_{\Sigma}\ ,
\end{equation}
where we extended slightly the original solution by allowing for an (near-horizon) extremal BTZ factor in the metric
instead of AdS$_3$,
\begin{equation}\label{btz}
 {\rm d} s^2_{BTZ} = \frac{1}{4} \left( - r^2\ {\rm d} t^2 + \frac{{\rm d} r^2}{r^2} \right) + \rho_+ \left( {\rm d} z + 
\left( -\frac{1}{4} +\frac{r}{2 \rho_+} \right) {\rm d} t \right)^2\ ,
\end{equation}
where $\rho_+$ is a constant that describes the mass of the BTZ black hole \cite{Banados:1992wn}.
Locally the BTZ and AdS$_3$ metrics are the same, therefore the equations of motion and BPS variations are not sensitive 
under this change. The solution for the gauge fields remains the same, \eqref{gaugefields}, while the scalars and BTZ 
and $\Sigma^2$ radii are given explicitly in terms of the magnetic charges,
\begin{equation}
 X^1 = \frac{p^1 (-p^1+p^2+p^3)}{(p^1 p^2 p^3 \Pi)^{1/3}}, \quad X^2 = \frac{p^2 (p^1-p^2+p^3)}{(p^1 p^2 p^3 
\Pi)^{1/3}}, \quad X^3 = \frac{p^3 (p^1+p^2-p^3)}{(p^1 p^2 p^3 \Pi)^{1/3}}\ ,
\end{equation}
\begin{equation}
R_{AdS_3}^3 = \frac{8 p^1 p^2 p^3 \Pi}{\Theta^3}\,, \qquad R_{\Sigma}^6 = \frac{(p^1 p^2 p^3)^2}{\Pi}\,,
\end{equation}
with
\begin{align}
 \Pi = &\, (p^1+p^2-p^3) (p^1-p^2+p^3) (-p^1+p^2+p^3)\,,
 \nonumber\\
\Theta =&\, 2 (p^1 p^2+p^1 p^3+p^2 p^3) - ((p^1)^2 + (p^2)^2 + (p^3)^2)\ .
\end{align}
Note that for the near-horizon geometry we actually needed to allow for a BTZ metric instead of AdS$_3$ in order to add 
the Wilson lines, since pure AdS$_3$ has a contractible cicle at the origin $r=0$. This is no longer true for the BTZ 
since the origin is covered by an event horizon and thus the Wilson line never shrinks outside the horizon. The full
asymptotically AdS$_5$ solution would then differ from the one in \cite{Benini:2013cda}, by the addition of the extra
charge described by $\rho_+$ and it is an interesting open problem to construct such a solution explicitly.

\section{Relation to axions in 4d} 
\label{sec:4d}

In this section we give a brief discussion of the analogous situation in four dimensions,
focusing on a theory with a known gauge theory dual and an M-theory lift. As already noted in section 3.6 of 
\cite{Hristov:2014eza}, the Wilson lines along the string in 5d correspond to axions in 4d. Here we show that adding electric charges to 
the BPS black holes of Cacciatori-Klemm (CK) \cite{Cacciatori:2009iz} in AdS$_4$ (see also \cite{Dall'Agata:2010gj,Hristov:2010ri,Halmagyi:2013qoa,Halmagyi:2013uza,Katmadas:2014faa,Halmagyi:2014qza} for more details and generalizations) is in one to one correspondence with switching on 
nonvanishing axions. This is the 4d analog of the Wilson line-electric charge correspondence in 5d and is known as the 
Witten effect \cite{Witten:1979ey}. 

Note that if we instead consider the Wilson lines along the Riemann surface, there is virtually no difference between the 4d and 5d solutions. The 4d gauged supergravity equations of motion and BPS variations are also invariant under the addition of flat connections \eqref{wilson} subject to\footnote{Note that in our 4d conventions the FI parameters are denoted $g_I$, therefore strictly speaking the new condition for BPS-ness of Wilson lines along the Riemann surface is
\begin{equation}\label{WilsonBPS4d}
 g_I w^I = 0\ .
\end{equation}
} \eqref{WilsonBPS}. Their dual interpretation is discussed in the next section. In the remaining discussion here we therefore focus on the case of CK black holes with axions.  

We consider an FI gauged theory in four dimensions with the prepotential
\begin{equation}\label{STU}
 F= \sqrt{X^0X^1X^2X^3}\,,
\end{equation}
and gauging 
\begin{equation}\label{eq:G-STU}
\cG=\left( 0, \squad 0, \squad 0, \squad 0, \squad g_1,\squad g_2, \squad g_3, \squad g_0 \right){}^{\!T}\,.
\end{equation}
In terms of the FI parameters, the AdS$_4$ radius and the scalars at the vacuum take the
form
\begin{gather}\label{eq:vac-STU}
R_{AdS} = I_4(G)^{-1/4}= (4 \, g_0 g_1 g_2 g_3)^{-1/4}\,, 
\qquad
t^i\bigr|_{\infty} =\mathrm{i}\,\sqrt{\frac{g_1g_2g_3}{g_0}}\,\frac{1}{g_i}\,,
\end{gather}
so that all axions vanish. It then follows that any nontrivial axions in a black hole
solution must vanish asymptotically and that the corresponding profile must be supported
by appropriate charges. Here, we used the notation $I_4$ for the quartic invariant of very
special geometry, given explicitly by
\begin{eqnarray}
I_4(\Gamma)
         &=& - (p^0 q_0 + p^i q_i)^2 + \frac{2}{3} \,p^0\,c_{ijk} p^i p^j p^k + \frac{2}{3} \,q_0\,c^{ijk} q_i q_j q_k 
             + c_{ijk}p^jp^k\,c^{ilm}q_lq_m\,. \label{I4-ch}
\end{eqnarray}
This invariant and its derivatives are crucial in the description of four dimensional black holes,
as will be made clear below. 

In order to explore axionic solutions, we will consider the attractor values for the scalars, which
are parametrised by the vector
\begin{equation} \label{eq:B-def}
 2\,\ee^{\psi-U}\I(\ee^{-\im\alpha}\cV)= \cB \,,
\end{equation}
where $U$ and $\psi$ parameterise the attractor geometry as
\begin{gather}\label{metric-fin}
  ds^2 = -\ee^{2U} r^2\,d t^2  + \ee^{-2U} \frac{dr^2}{r^2} 
  + \ee^{2(\psi-U)}\, \left( d\theta^2 + \sin^2{\theta} d\phi^2 \right)\,,
\end{gather}
The attractor equations that determine the scalars in terms of the gauging, $G$, and
the charge vector, $\Gamma$, are given by \cite{Katmadas:2014faa}
\begin{equation}\label{eq:Gam-BB}
 \tfrac14\, I^\prime_4(\cB, \cB, \cG ) = \Gamma\,,
\end{equation}
along with the constraints
\begin{equation}\label{eq:constr-4d}
 \Iprod{\cG}{\Gamma}=-1\,, \qquad \Iprod{\cB}{\Gamma} = 0\,.
\end{equation}

In order to show the correspondence between electric charges and axions, we introduce
the following decomposition of the $2\, \nv +2 $ dimensional vector space in four
subspaces, as
\begin{equation}\label{eq:decomp}
 \left(\begin{array}{c} 
p^0 \\ p^i \\ q_i \\ q_0 
\end{array}\right)
 \begin{array}{c} (+3) \\ (-1) \\ (+1)  \\ (-3) \end{array}\,,
\end{equation}
which assigns one of the eigenvalues $\pm 3$ and $\pm 1$ to the various components of
any symplectic vector. The details of this decomposition and the concrete operators that
project to the various subspaces can be found in \cite{Bossard:2013oga}.

Clearly, the decomposition \eqref{eq:decomp} is consistent with the quartic invariant \eqref{I4-ch}, since
each term appearing in that expression is manifestly of total grade zero. This holds for
any combination $I_4(\Gamma_1,\Gamma_2,\Gamma_3,\Gamma_4)$, which can be nonzero only if the
various vectors have appropriate components so that a grade zero term can be constructed.
Similarly, any derivative $I^\prime_4(\Gamma_1,\Gamma_2,\Gamma_3)$ is nonzero only if
components of eigenvalue $\pm 3$ and $\pm 1$ can be constructed.

This decomposition is particularly relevant for the gauging \eqref{eq:G-STU}, which lives
in the $(-3)\oplus(+1)$ subspace. For nontrivial axions to appear in this basis, the vector
$\cB$ must have components in both the $(\pm 1)$ subspaces, as is clear from the expressions
for the scalars at infinity \eqref{eq:vac-STU} and can be verified directly from the explicit
solution to \eqref{eq:B-def} for the physical dilatons and axions, see e.g.\ \cite{Bates:2003vx}.

As shown recently in \cite{Halmagyi:2014qza}, the general asymptotically AdS$_4$ black hole
solution can be parametrised in terms of four vectors constructed out of the gauging and charge
vectors. In particular, we are interested in the attractor solution for $\cB$, which reads
\begin{equation}\label{att-sol-4d}
 \cB = a_1 I^\prime_4(\cG,\cG,\cG) + a_2 I^\prime_4(\cG,\cG,\Gamma) 
      + a_3 I^\prime_4(\cG,\Gamma,\Gamma) + a_4 I^\prime_4(\Gamma,\Gamma,\Gamma)\,,
\end{equation}
where the constant coefficients $a_i$ are given in terms of the gaugings and charges by
\eqref{eq:ai}.

Given that the first vector in \eqref{att-sol-4d}, $I^\prime_4(\cG,\cG,\cG)$ is by assumption
of grade $(-1)\oplus(+3)$ due to \eqref{eq:G-STU}, an axion free solution at the attractor can
only be realised if the other three vectors are of the same type. It is straightforward to verify that
this is only possible for $\Iprod{\Gamma}{G}\neq 0$ if $\Gamma$ is of grade $(-1)\oplus(+3)$,
i.e.\ purely magnetic according to \eqref{eq:decomp}, and $a_2=a_4=0$. In this case, the second
constraint in \eqref{eq:constr-4d} is trivially satisfied. 

Turning on the remaining, electric, components of $\Gamma$ along the directions $(-3)\oplus(+1)$
leads to all possible directions turned on in \eqref{att-sol-4d}, while \eqref{eq:constr-4d}
becomes nontrivial. Using the latter to fix the eigenvalue $(-3)$ component in \eqref{att-sol-4d},
we find that the $\nv$ components of degree $(+1)$ in the charge, i.e.\ the electric charges $q_i$,
are necessary and sufficient to turn on the axions in $\cB$.

\section{Dual field theory interpretation} 
\label{sec:cft}

\subsection{Wilson lines along the string/axions}
\label{4.1}
It is interesting to observe that the addition of Wilson lines along the string in the 5d bulk leads to a nontrivial winding around the circle of the scalars in the boundary $N=4$ SYM. The same phenomenon was observed in the MSW theory \cite{Maldacena:1997de} upon addition of membrane charges (which are again Wilson lines from the black string perspective) and the situation in our case is in complete analogy. Consider the twisted $N=4$ SYM, dual to the BB black strings. There is a non-vanishing background gauge field for the R-symmetry, so the fields of $N=4$ SYM feel the covariant derivative 
\begin{equation}
 D_{\mu} = \partial_{\mu} + i s \tilde{\omega}_{\mu} - i A_{\mu}^B\ ,
\end{equation}
where $s$ corresponds to the spin of the field and $A^B$ is a background $U(1)^3 \subset SO(6)_R$ gauge field in the Cartan subgroup that describes the magnetic flux through the Riemann surface $\Sigma$ (see \cite{Benini:2013cda} for detailed explanation). The addition of the Wilson line means that there is now also a constant nonvanishing $A_{\gamma}^B$ on the circle of the boundary. This is a background Wilson line, and is given by a linear combination of the values $w^I$ of the bulk Wilson lines \eqref{wilson}. In the conceptually simplest case of scalar fields (no spin connection in the covariant derivative), the scalars can no longer be given by constant modes on S$^1$, but instead acquire winding numbers around the circle to cancel the background Wilson line.

Note that the black string metric remains unchanged upon the addition of Wilson lines, and therefore the 
main features of the RG flow of the twisted $N=4$ SYM also remain unchanged. In particular, the central charge of the 
final $N=(0,2)$ 2d SCFT remains the same, i.e.\ the number of multiplets in the IR theory is unchanged. This is not 
surprising as the addition of the background Wilson line on the boundary is not related to the coupling of the original theory to the Riemann surface that is eventually relevant for counting the central charge \cite{Benini:2013cda}. The Wilson lines along the string are completely decoupled from physics related to the internal space, so one might even think of them directly in 3d context after compactifying the 5d supergravity on the Riemann surface \cite{Karndumri:2013iqa}. The situation would then be described via the AdS$_3$/CFT$_2$ correspondence, see e.g.\ \cite{Gupta:2008ki} for details.

We can also look more qualitatively on the dual field theory interpretation of our results and make comparison between the field theory duals of 5d and 4d asymptotically AdS solutions carrying magnetic charge. Switching on nonvanishing expectation values for Wilson lines in the bulk translates into switching on an extra deformation on the boundary. The exact same type of logic holds in the 4d bulk, where switching on axions leads to an extra deformation related to the mass of the fermions in the dual theory. In one case we are looking at the RG flow of twisted 4d $N=4$ SYM to a 2d $N=(0,2)$ SCFT, while in the other case we have the RG flow of twisted 3d ABJM theory to 
a $N=2$ superconformal quantum mechanics\footnote{Strictly speaking the electric charges mean that we are looking at excited states that no longer preserve the full superconformal symmetry. Here we refer to the underlying 2d/1d superconformal theories, which correspond to the solutions not yet deformed by Wilson lines/axions.}. Even though the dimensions do not match, a similar phenomenon occurs, 
that seems to be related to the fact that both theories are twisted. In both cases the extra deformations (from Wilson lines 
or axions) lead to the switch of the BPS states of the theory from a zero to a nonzero charge. This is the case because 
the electric charges in the bulk translate into $U(1)$ charge densities of states on the boundary \cite{Gupta:2008ki}. The fact that in the bulk solutions Wilson lines/axions go hand in hand with electric charges was shown to be related to the existence of magnetic charges, without which no electric charges would exist. Magnetic charges lead to a twisting of the boundary theory as explained in \cite{Benini:2013cda,Hristov:2013spa}. Therefore we observe a new phenomenon on the dual field theory side: due to the twisting, BPS states in the theory acquire $U(1)$ charge densities upon the switch of the extra BPS deformation from Wilson lines/axions. 

\subsection{Wilson lines along the Riemann surface}
\label{4.2}
Putting the Wilson lines on the Riemann surface both in 4d and in 5d seems to be in complete analogy to the situation in 7d, already discussed in some detail in section 4.3 of \cite{Bah:2012dg}. Here we just summarize briefly their main conclusion, which is that the Wilson lines in the bulk correspond to a set of exactly marginal deformations of the boundary field theory. Upon compactification of the $N=4$ SYM or the ABJM theory on a Riemann surface, the resulting 2d/1d superconformal theory is expected to have a space of exactly marginal deformations. This space is locally a product of the complex structure deformations of the Riemann surface and the space of Wilson lines for abelian global symmetries in the theory. Equation \eqref{WilsonBPS} (or equivalently \eqref{WilsonBPS4d}) tells us that we cannot turn on a Wilson line for the R-symmetry, therefore the number of Wilson lines we can switch on for $N=4$ SYM is 2, and for ABJM 3. For a Riemann surface with a genus $g$, the space of flat connections has a complex dimension $g$, therefore we expect the total dimension of the space of marginal deformations for Wilson lines to be $2 g$ and $3 g$ for the theories coming from compactification of $N=4$ SYM and ABJM, respectively. Finally note that marginal deformations keep the central charge invariant, in accordance with the fact that the metric is unaffected by the addition of Wilson lines.

\section*{Acknowledgments}
We would like to thank N.\ Bobev and A.\ Zaffaroni for enlightening discussions. We are supported in part by INFN, by the MIUR-FIRB 
grant RBFR10QS5J ``String Theory and Fundamental Interactions'', and by the MIUR-PRIN contract 2009-KHZKRX.

\begin{appendix}
 \section{Explicit form of the attractor solution in 4D}
 
In this appendix we discuss briefly the explicit solution to the attractor equation \eqref{eq:Gam-BB},
following the method of \cite{Halmagyi:2014qza}. For the convenience of the reader, we first display the
explicit form of the left hand side of \eqref{eq:Gam-BB} for the gauging \eqref{eq:G-STU}
and the components of the vector $\cB$ defined as
\begin{equation}
\cB = \left( \beta^0\,, \quad \beta^i \,, \quad \beta_i \,, \quad \beta_0 \right){}^{\!T}\,,
\end{equation}
With this notation, the left hand side of \eqref{eq:Gam-BB} reads
\begin{equation}\label{eq:I4BBG-ex}
\tfrac14\,I^\prime_4(\cB,\cB,\cG) = 
\begin{pmatrix}
\beta^0\, g^I\beta_I  + c^{ijk}\,g_i \beta_j \beta_k 
\\ 
\\ 
- \beta^i\, g^I\beta_I   + 2\,\beta^i \sum_{ j \neq i}\beta^jg_j 
 + c^{ijk}( 2\,\beta_0 \beta_j g_k + g_0 \beta_j \beta_k )
\\ 
\\ 
g_i\, \beta^I\beta_I + \beta_i\, g^I\beta_I  - 2\,\beta_i \sum_{ j \neq i}\beta^jg_j 
 - 2\,g_i \sum_{ j \neq i}\beta^j\beta_j 
\\ 
g_0 \, \beta^I\beta_I + \beta_0\, g^I\beta_I
\\ 
\end{pmatrix}\,,
\end{equation}
where we used the shorthand notation
\begin{align}
\beta^I\beta_I = \beta^0 \beta_0 + \beta^i \beta_i\,,
\qquad
g^I\beta_I  = g_0 \beta^0 + g_i \beta^i \,,
\end{align}
and $c^{ijk}$ is the completely symmetric tensor which is $c^{123}=1$ and and vanishes
when any two indices are equal.

We will not show the solution for $\cB$ in components, but rather use the basis in \eqref{att-sol-4d}
to express the full vector and give the general duality covariant solution for any charge and
gauging, by specifying the constants $a_i$ as
\begin{align}\label{eq:ai}
 a_1 = &\, 
 6\,\frac{I_4(\cG,\cG,\cG,\Gamma)^2 I_4(\Gamma) - I_4(\cG,\Gamma,\Gamma,\Gamma)^2 I_4(\cG) 
}{I_4(\cG,\cG,\cG,\Gamma)^3}\,a_4 \,,
\nonumber
\\
 a_2 = &\, -\frac{I_4(\cG,\Gamma,\Gamma,\Gamma)}{2\,I_4(\cG,\cG,\cG,\Gamma)}\,a_4\,, 
\nonumber
\\
 a_3 = &\, -3\,\frac{I_4(\cG,\cG,\cG,\Gamma)^2 I_4(\Gamma) - I_4(\cG,\Gamma,\Gamma,\Gamma)^2 I_4(\cG) 
}{I_4(\cG,\Gamma,\Gamma,\Gamma) I_4(\cG,\cG,\cG,\Gamma)^2}\,a_4 \,,
\nonumber
\\
 a_4 = &\, \frac{I_4(\cG,\cG,\cG,\Gamma)^2 I_4(\cG,\Gamma,\Gamma,\Gamma)}{ 
 \sqrt{ b_1\, b_2 } }
            \,,
\\
b_1 = &\,-2\, ( \Iprod{\Gamma}{\cG}\,(\cG,\cG,\Gamma,\Gamma) + \Iprod{I^\prime_4(\cG)}{I^\prime_4(\Gamma)} ) \,,
\nonumber
\\
b_2 = &\,I_4(\cG,\cG,\cG,\Gamma)^3 I_4(\cG,\Gamma,\Gamma,\Gamma)^3 +36\,(I_4(\cG,\cG,\cG,\Gamma)^2 I_4(\Gamma) - 
I_4(\cG,\Gamma,\Gamma,\Gamma)^2 I_4(\cG))^2 
\,.
\nonumber
\end{align}
We stress that these expressions differ by the overall normalization, through $a_4$, from the attractor
solution one would obtain by taking a limit of the solution given in \cite{Halmagyi:2014qza} as we are using
the conventions of \cite{Katmadas:2014faa} for the attractor equations.

\end{appendix}

\providecommand{\href}[2]{#2}
\end{document}